\documentstyle[11pt,aaspp4,psfig]{article}
\input amssym.def
\input amssym 

\newcommand{\kms}	{km~s$^{-1}$}
\newcommand{\h}   	{$h^{-1}\,$~kpc}
\newcommand{\etal} 	{{et~al.}}

\newcommand{\pcm}	{cm$^{-2}$}

\newcommand{\hst}	{{\it HST}}

\received{July 17 1996}
\accepted{September 17 1996}

\begin{document}

\title{
Complex Mg~II absorption in the outer disk of M61$^1$}

\author{David V. Bowen$^{2}$, J. Chris Blades$^3$, and Max Pettini$^4$}

\affil{ $\:$}

\affil{$^2$Royal Observatory, Edinburgh, Blackford
Hill, Edinburgh EH9 3HJ, United Kingdom\\email: dvb@roe.ac.uk}

\affil{$^3$Space Telescope Science Institute, 
3700 San Martin Drive, Baltimore, MD 21218\\email: blades@stsci.edu }

\affil{$^4$Royal Greenwich Observatory, Madingley Rd., Cambridge CB3 0EZ
\\email: pettini@ast.cam.ac.uk}

\altaffiltext{1}{Based on observations obtained with the NASA/ESA Hubble 
Space Telescope, obtained at STScI, which is operated by the Association of
Universities for Research in Astronomy, Inc., under contract with the National
Aeronautics and Space Administration, NAS5-26555.}

\begin{abstract}

The increasing availability of high quality spectra of QSO
absorption line systems at resolutions of only a few \kms\ 
is expected to facilitate the translation of the
kinematics of components comprising the lines into the spatial distribution of
gas around an absorbing galaxy. In this {\it Letter}, we present {\it Hubble
Space Telescope} ({\it HST}) spectra of Q1219+047, a QSO whose sightline
passes 21~\h\ from the center of M61, through the outer regions of an extended
H~I disk.  We detect complex Mg~II absorption, spanning a velocity range of
$\approx 300$~\kms , and strong C~IV absorption; these are the first UV
observations of a QSO absorption line system arising in the outskirts of a disk
of a nearby galaxy at low inclination.  Our observations are at odds with
models of galaxies in which absorbing clouds co-rotate with a galaxy's disk,
because M61's low inclination should give rise to only a few Mg~II 
components spread over a small velocity range in such a model. Hence
our results throw doubt on whether absorption line profiles can be used to
infer the spatial distribution of gas around absorbing galaxies.

\end{abstract}


\section{Introduction}

It has long been hoped that the distribution of subcomponents in QSO
absorption line complexes would prove instrumental in understanding how gas is
distributed around galaxies.  For example, the asymmetric profiles of low
ionization species observed with the {\it Keck Telescope} in high redshift
damped Ly$\alpha$ systems are interpreted as being due to the line of sight to
the background QSO passing through a rotating disk (Wolfe 1995; Prochaska \&
Wolfe 1996), in keeping with the hypothesis that such systems are the early
progenitors of late type galaxies.  The direct detection of galaxies
responsible for Mg~II absorption systems at redshifts $0.2 < z < 1$ 
has provided an opportunity to examine whether the
profiles of the absorption lines are related to the properties of the
galaxies, particularly their distance from the QSO sightline (`impact
parameter'), their luminosity, or their inclination.  From the small sample of
absorbing galaxies surveyed by 
Bergeron \& Boiss\'{e} (1991), Lanzetta \&
Bowen (1992) used high-resolution (FWHM of $7-22$~\kms ) 
data of the Mg~II lines to
propose that the variation in strength of the sub-components with velocity was
consistent with that expected for absorption by a rotating ensemble of clouds,
and that the complexity should decrease with impact parameter.  In their
model, the important parameter determining the variation in the sub-structure
along many different lines of sight is the inclination of the absorbing disk.

Unfortunately, the identification of a larger sample of absorbing galaxies
(Steidel 1995; Steidel, Dickinson \& Persson 1994) and high resolution spectra
(Churchill, Steidel \& Vogt 1996) has failed to confirm that the distribution
of sub-components within a line depends on impact parameter.  Observations of
{\it nearby} galaxies also show that the characteristics of Mg~II absorption
lines depend on a wide variety of physical mechanisms (particularly whether
the galaxy is interacting with neighbouring galaxies), and not simply on
impact parameter or galaxy inclination (Bowen, Blades \& Pettini 1995a,
hereafter BBP).  Moreover, in our own Galaxy, Ca~II and Na~I absorption lines
observed toward distant stars or extragalactic sources, (which arise in high
H~I column densities and which should therefore trace the bulk of the disk
gas) are more often found at velocities which are inconsistent with gas
co-rotating with the disk when observed at high resolution 
(Morton \& Blades 1986;
Sembach, Danks \& Savage 1993).


In this {\it Letter}, we present observations of Q1219+047, a QSO whose line
of sight passes 4.67$'$ from the center of M61 ($\equiv$ NGC~4303 ; see
Fig.~\ref{fig:grey}), which corresponds to 21.1~\h\ at the galaxy's velocity
of 1569~\kms .  
M61 is a grand-design spiral [of type SAB(rs)bc] located near the outer edge
of the Virgo Cluster at a projected distance of 8.2 degrees from M87. 
It shows little H~I deficiency (Giovanelli \& Haynes 1985) but may have a high
star formation rate (Kennicutt 1983).
M61 is a galaxy at low inclination ($i = 25^o$), and the QSO
sightline intercepts the disk at a distance of 1.5 times the optical radius
(measured to a surface brightness of 25 mag arcsec$^{-2}$).  With the spectra
presented in this paper,
we test explicitly whether the absorbing gas in the outer
regions of a normal spiral galaxy follows the velocity field of the bulk of the
disk gas. If it does, we would expect absorption lines to have a simple
structure because of the galaxy's low inclination, causing the line of sight
velocity component to be small. In fact, we find that the Mg~II absorption
along the line of sight is clearly complex and contains components
spanning a total velocity range of nearly 300~\kms .

\section{Observations and Results}

Q1219+047 was observed with the Faint Object Spectrograph (FOS) and 1.0 arcsec
circular aperture (B-3) using the G130H and G270H gratings.  The QSO ($z =
0.094$) has a $V$-band magnitude of 16.8, and was observed for 34.0 mins with
the G270H and 38.3 mins with the G130H. The spectra were calibrated using
standard pipeline software.  To obtain a more exact zero point for the
wavelength calibration of the G270H data, we compared the velocity of
five strong Fe~II absorption lines from Milky Way gas in
the QSO spectra, with H~I emission along the QSO line of sight taken from the
Leiden/Dwingeloo 21~cm H~I survey (Hartmann 1994).  A shift of $\approx
+1.5$~\AA\ ($\equiv 160$~\kms ) was required to match the velocities. The
G130H spectrum was of too low a signal to noise to measure accurately the
wavelengths of Milky Way absorption lines and improve its calibration.

Both G130H and G270H spectra are presented in full, along with complete line
identifications, in a forthcoming paper.  Plots of the normalised spectra
around the position of Mg~II and C~IV are shown in Figure~\ref{fig:spec}. The
signal-to-noise of these data are 4 and 13 per pixel for the G130H and G270H
gratings, respectively. The figure marks the rest wavelengths of
absorption expected from our own Galaxy, as well as the wavelengths of
absorption expected at the systemic velocity of M61.  
Strong Mg~II is detected from both the Milky Way and from M61:  the equivalent
widths, $W$, of the lines observed from M61 are $W$(Mg~II$\lambda 2796$)$ =
1.96\pm 0.16$~\AA, $W$(Mg~II$\lambda 2803$)$ = 2.00\pm 0.18$~\AA , $W$(C~IV$
\lambda 1548$)$ = 1.28\pm 0.25$~\AA , and $W$(C~IV$ \lambda 1550$)$ = 0.63\pm
0.25$~\AA .  From our own galaxy, we find local absorption at strengths of
$W$(Mg~II$\lambda 2796$)$ = 0.87\pm 0.13$~\AA, and $W$(Mg~II$\lambda 2803$)$ =
1.02\pm 0.13$~\AA .  In the lower signal-to-noise G130 spectrum, C~IV is
absent from Milky Way gas.

A 21~cm H~I map of M61 has been constructed by Cayatte~et~al.\ (1990).  The
sightline to Q1219+047 passes just beyond the last H~I column density contour
of $10^{20}$~cm$^{-2}$, which we indicate by the dashed ellipse in
Fig.~\ref{fig:grey}.  This value is approximately 1$-$1/8 th times the column
density 
we would expect to see from our own galaxy observed face-on (e.g. Elvis,
Lockman \& Wilkes 1989; Diplas \& Savage 1994), but since the Mg~II$\lambda\lambda 2796,2803$
absorption from the Milky Way is already saturated (e.g. Bowen, Blades \&
Pettini 1995b), we would predict that the
absorption from M61 would be of similar strength to Galactic absorption. The
fact that it is stronger means that the velocity spread of components
comprising the line is larger than that seen locally.
In particular, as can be seen in Fig.~\ref{fig:spec}, 
components of the Mg~II are clearly resolved blueward
of the bulk of the absorption.
Since the FOS has a resolution of $\simeq 200$~\kms\ FWHM for point
sources at this wavelength 
the Mg~II line must be composed of components which span at least this
velocity range. To illustrate this point, Fig.~\ref{fig:spec} shows
theoretical absorption line Voigt profiles convolved with the Line Spread
Function (LSF) of the FOS, assuming that the LSF is gaussian and 0.96 diodes
FWHM. We have chosen column densities, $N$, and Doppler parameters, $b$, such
that the instrumental LSF dominates the shape of the profile ($b < 40$~\kms
). 
The dotted line shows the shape of the profile for
a single absorbing cloud, equivalent to many smaller clouds (with similar $b$
values) spread over a velocity range less than the FWHM of the LSF. The excess
absorption in the blue wing of both Mg~II lines 
can be clearly seen. To estimate the velocity range over which this
excess absorption extends, we have simply fit a second single cloud to the
excess absorption. We find that additional absorption is required at
velocities $\approx -290$~\kms\ different from the bulk of the absorption.

Unfortunately, the G130H spectra is of too low a signal-to-noise to reveal any
excess absorption in the C~IV absorption lines from M61. 
Nevertheless, it is clearly
strong compared to Milky Way absorption which is not detected.  Unlike Mg~II,
C~IV absorption lines in our own halo are not saturated, and the strong lines
seen in M61 probably reflect a significantly larger column density than that
detected locally. Considering that the H~I column density is lower along the
line of sight, this suggests that some portion of the gas in the disk and/or
halo of M61 is more highly ionized than it is in the Milky Way. Discerning
whether this highly ionized gas is found over the same velocity range as the
Mg~II lines, or whether it arises exclusively in the same components marked by
the excess Mg~II absorption at $\approx -290$~\kms\ blueward of the bulk of
the absorption, will require higher signal-to-noise data.

\section{Discussion}

Since M61 is at low inclination, we would expect any velocity component from
absorbing gas along our line of sight originating 
from the disk of the galaxy to be
small. However, the Mg~II absorption is clearly complex and spans a velocity
range $\Delta v \simeq 300$~\kms . This is larger than the range of velocities
seen from the H~I gas observed across the entire galaxy ($\Delta v
\simeq 200$~\kms ); hence the complexity of the absorption cannot be related
to the inclination of M61's disk to the line of sight.

The bulk of the Mg~II absorption occurs at $v_\odot \simeq 1560$~\kms , close
to the systemic velocity of the galaxy (1569~\kms ). However, the velocity of
the H~I gas seen in 21~cm emission near the sightline is $\simeq 1640$~\kms ,
some 80~\kms\ higher than the Mg~II absorption.  As noted in \S2, we have
attempted to zero-point the wavelength calibration of the spectrum as
accurately as possible.  80~\kms\ corresponds to a shift of only 1.5 pixels,
so we cannot be sure that the measurement of the bulk of the absorption is in
error by a similar amount.  H~I gas is detected at 21~cm from column densities
which can be several dex larger than those in which the Mg~II occurs (Mg~II
absorption lines of the strength detected here 
arise in gas which is optically thick at the Lyman Limit, i.e., 
$N$(H~I)$> 2\times 10^{17}$~\pcm ); hence Mg~II absorption
lines will arise from the same H~I detected at 21~cm, but could, in addition,
arise from gas which remains undetected at 21~cm. Thus the observed Mg~II
absorption does not have to be centered 
at the same velocity as the H~I.  Whether the
difference in velocity is real, or due to wavelength calibration errors, will
require further observations, but the uncertainty in the absolute velocity of
the absorption does not affect the {\it relative} velocity difference between
the excess absorption and the main body of the Mg~II line.

The large velocity range over which Mg~II absorption is observed
suggests that the distribution of sub-components
comprising the absorption complex cannot be related to the velocity field of
the disk of M61. Asymmetric profiles similar in shape to that observed in M61
are predicted for highly inclined disks (Lanzetta \& Bowen 1992), not for
galaxies at low inclination.  
Our observations therefore show that the shape of absorption line profiles
seen in higher-redshift absorption systems need not have anything to do with
rotating ensembles of gas clouds in inclined galaxy disks.

The complicated absorption could be evidence of the two component model for
Mg~II absorption, in which Mg~II absorption arises in both high $N$(H~I) disk
gas, and lower $N$(H~I) halo gas (Briggs \& Wolfe 1983).  The bulk of the
Mg~II absorption would be expected to arise in the disk of the galaxy in this
model, and the fact that it does not appear to be close to the velocity of the
H~I gas in the disk measured from 21~cm emission is hard to
understand. Nevertheless, as we have noted, the absolute wavelength scale of
the FOS spectrum may be in error. Even disregarding this observation, the {\it
relative} velocity difference of $\approx 300$~\kms\ between disk and halo gas
is just as hard to interpret.
If the halo were completely at rest with respect to the disk of the
galaxy, the difference in velocity between halo and disk gas along the line of
sight should be no more than the maximum amplitude 
of the galaxy's rotation curve.
Again, the
21~cm maps of M61 show that this value should be $\approx
100$~\kms . Hence we believe that a simple two component disk-halo model of
the absorbing gas may not be adequate to describe the observed absorption.


It is also possible that the complex absorption has nothing to do with gas
intrinsic to M61, but is the superposition of absorption lines from the
overlapping halos of nearby galaxies. There are two nearby galaxies, NGC~4292
and NGC~4303A, 7.58$'$ and 12.66$'$ away from the QSO sightline,
respectively. The velocity of NGC~4292 is $cz_{\rm{gal}} = 2258$~\kms ,
690~\kms\ higher than M61, which puts it $\simeq 1000$~\kms\ away from the
velocity of the excess absorption seen in the QSO spectrum. For this reason
alone, it seems unlikely that NGC~4292 contributes to the absorption complex,
as there is no evidence for such large motions in the halos of normal
galaxies.  NGC~4303A is closer in velocity to M61, $cz_{\rm{gal}} = 1273$~\kms
, a difference of $- 290$~\kms\ between the galaxy and M61, almost the same as
that detected in the QSO spectrum.  In this model where single galaxies are
surrounded by spherical halos, the radius of the absorbing cross-section,
$R_{\rm{Mg~II}}$, is related to the galaxy's luminosity, $L$, by
$R_{\rm{Mg~II}} \approx 35 h^{-1} (L/L^*)^{0.2}$ (Steidel 1995), at least for
galaxies at redshifts $0.2 < z < 1.0$.  In Table~1 we list the separation of
the galaxies from the QSO line of sight, $\rho$, and the value of
$R_{\rm{Mg~II}}$ (for $M^* = -19.5$). For NGC~4292 \& NGC~4303A, $\rho /
R_{\rm{Mg~II}} > 1$, that is, the predicted radii of the spherical Mg~II
halos simply do not extend far enough to intercept the QSO line of sight.
It therefore seems unlikely that the complex absorption arises from the
overlapping halos of neighbouring galaxies.

If the complex absorption seen toward M61 does not arise from systematic---as
opposed to random---gas motions within a galaxy, what else might cause the
complex absorption? The large velocity range which the absorption covers is
most reminiscent of absorption from the High Velocity Clouds (HVCs) seen in
our Galaxy (Savage~et~al.\ 1993; Bowen, Blades \& Pettini 1995b;
Sembach~et~al.\ 1995 and refs therein). The problem in relating Galactic HVCs
to the more general phenomenon of QSO absorption lines is that their origin is
still unresolved. There is evidence of HVCs in some external galaxies, such as
M101 and NGC~628 (van der Hulst \& Sancisi 1988; Kamphuis 1993), which are
both galaxies with small inclinations and which show H~I complexes with sizes
of several kpc. In both cases, the difference in velocity between the HVC
complexes and the systemic velocity of the galaxy are only of order 100~\kms ,
less than the 300~\kms\ found for the Mg~II complex in M61. As noted
above though, the HVCs
and the Mg~II lines arise in H~I column densities which
can differ by several dex, and they may therefore have completely 
different velocity
structures. 

One of the explanations for the origins of some of the Milky Way HVCs is that
they are the result of interactions between the Galaxy and neighbouring
satellites. Since M61 has the two close neighbours discussed above, it is
possible that such interactions may have influenced the distribution of H~I in
the galaxy, both spatially and kinematically. There is no evidence for the
more severe distributions which lead to large, extended tidal tails spread
well away from the center of a galaxy, and which can certainly lead to the
observations of complex metal line absorption (BBP, Bowen~\etal\ 1994),
although the high rate of supernovae and the high density of H~II regions
(Hodge \& Kennicutt 1983) seen in M61 could have been triggered by such
interactions.

The effects of galaxy-galaxy interactions on the properties of QSO absorption
line systems have been largely ignored, partly because there are too few of
the `classic' interacting systems seen at the present epoch to account for the
number of systems detected at higher-redshift, but also because it is
difficult to generalize what the effects of interactions would be on the
observed absorption systems. In the case of M61, the effect is even more
subtle, if complex Mg~II absorption has been induced from very mild
galaxy-galaxy interactions which do not show up in any other obvious way. If
the interaction rate between galaxies has been higher in the past, as is being
inferred from \hst\ images of the distant universe (Abraham~\etal\ 1996; van
den Bergh~\etal\ 1996) then the effects of interactions on even apparently
normal galaxies should be evident in the characteristics of the absorption
lines which arise from gas associated with those galaxies.

Whether the complex absorption seen in the disk of M61 arises because of
isolated HVCs in the galaxies, from weak galaxy-galaxy interactions, or simply
from normal interstellar processes, the line profiles indicate that it may not
be so easy to translate the kinematic structure of high resolution absorption
line profiles to the spatial structure of interstellar gas in absorbing
galaxies.  Future observations with {\it HST} and STIS (particularly of the
higher ionization lines) will enable us to
confirm our result and probe the kinematics of M61's disk in more detail.


\bigskip

We thank Dap Hartmann for providing H~I spectra
from the Dwingeloo/Leiden 21~cm survey to help calibrate the FOS data.
DVB and JCB acknowledge financial support from HST 
grant GO-5451.


\clearpage
\begin{deluxetable}{lccccc}
\tablenum{1}
\tablewidth{0pc}
\tablecaption{GALAXIES CLOSE TO THE SIGHTLINE OF Q1219+047}
\tablehead{
\colhead{} 
& \colhead{$v_{\rm{gal}}$}
& \colhead{}
& \colhead{$\rho$}
& \colhead{$R_{\rm{Mg~II}}$} 
& \colhead{} \nl
\colhead{Galaxy} 
& \colhead{(\kms )}
& \colhead{$M_B$}
& \colhead{(\h )}
& \colhead{(\h )}
& \colhead{$\rho / R_{\rm{Mg~II}}$}
}
\startdata
M61		& 1641	& $-20.9$	& 21	& 45	& 0.5 \nl
NGC~4292	& 2258	& $-18.7$	& 49	& 30	& 1.6 \nl
NGC~4303A	& 1273	& $-17.1$	& 47	& 23	& 2.0 \nl
\enddata
\end{deluxetable}


\clearpage

\figcaption{Image of M61 (NGC~4303) reproduced from the STScI
Digitised Sky Survey.  The position of the QSO is marked. The sightline passes
just beyond the last H~I contour plotted by Cayatte et al.\ (1990), and
represented here by a dashed ellipse. \label{fig:grey} }

\figcaption{Portions of the normalised FOS spectra of Q1219+047
around the wavelength region expected for Mg~II (upper panel) and C~IV (lower
panel) absorption from M61.  Unresolved Mg~II absorption lines are seen from
our own Milky Way, but the Mg~II detected from M61 is clearly resolved and
contains an asymmetric blue wing. Theoretical line profiles are shown to
illustrate the LSF of the FOS (dotted line) and the extra
component 290~\kms\ to the blue. Strong C~IV absorption is also detected in
M61, whereas C~IV from our own Galaxy is absent.
\label{fig:spec}
}


\setcounter{figure}{0}

\clearpage
\begin{figure}
\centerline{\psfig
{figure=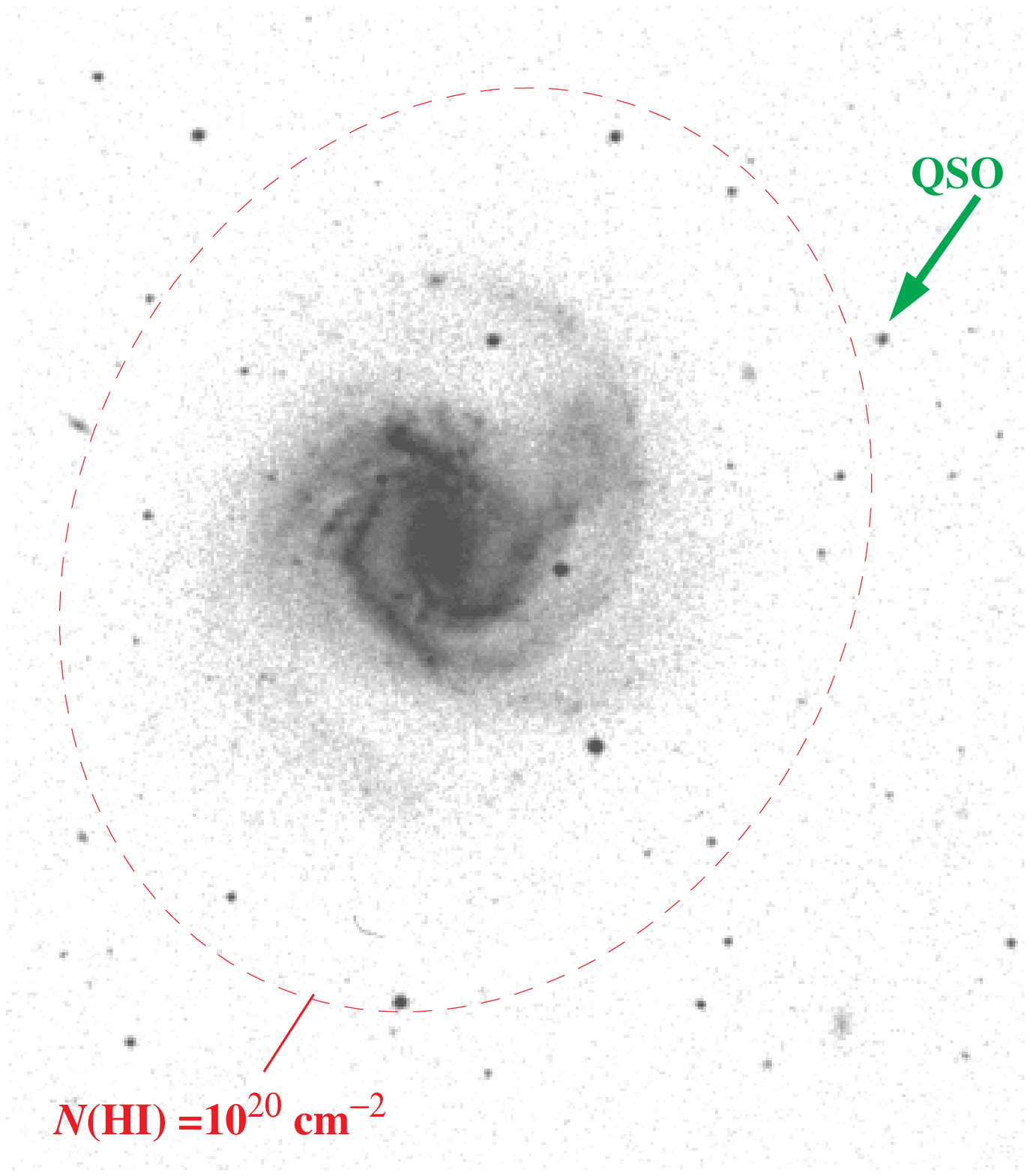,height=25cm}}
\caption{\label{fig:grey}}
\end{figure}

\clearpage
\begin{figure}
\centerline{\psfig
{figure=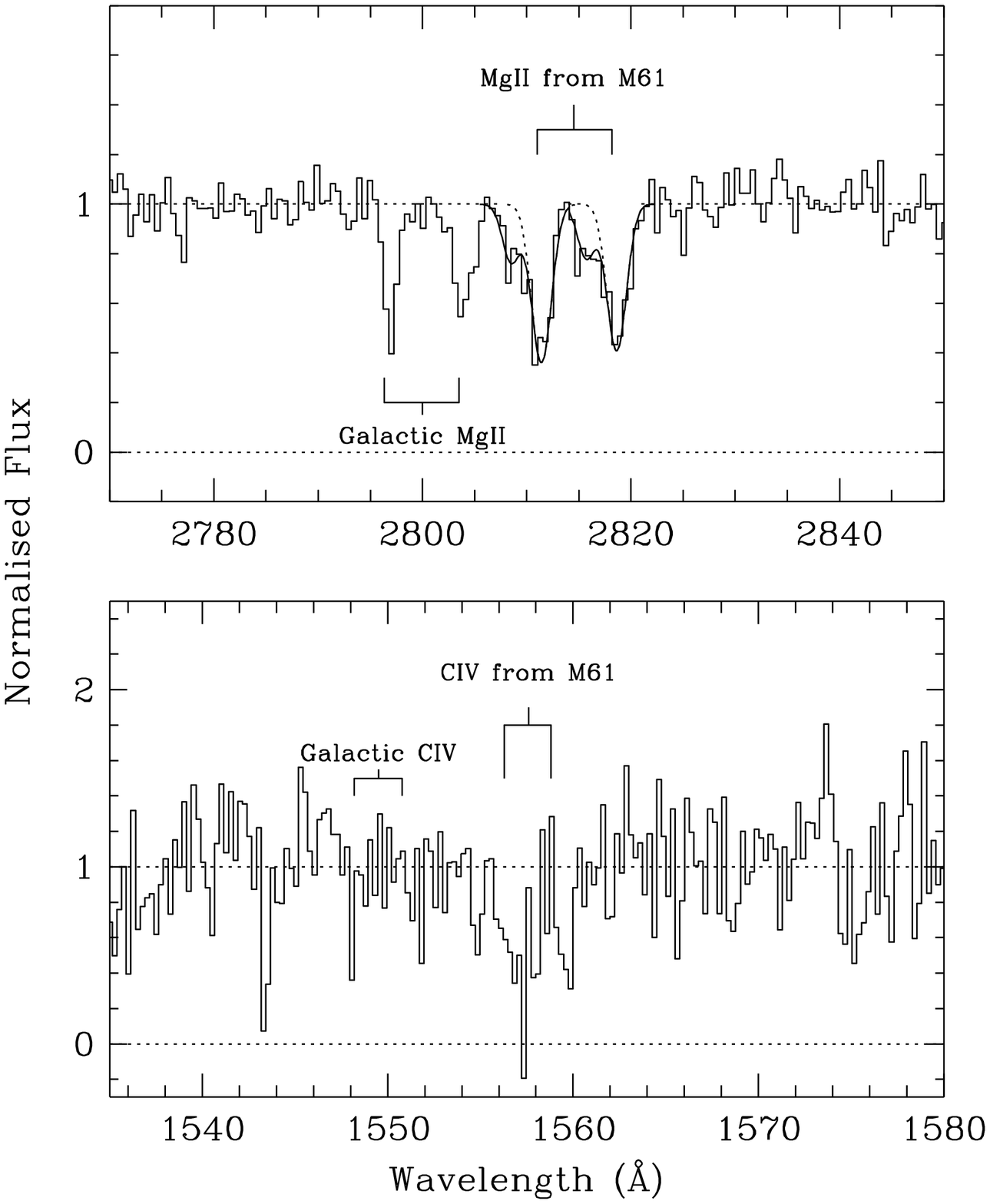,height=23cm}}
\caption{\label{fig:spec}}
\end{figure}


\end{document}